\documentclass[aps,prd,amssymb,nofootinbib,twocolumn,epsf,floatfix]{revtex4}
\usepackage[usenames]{color}
\usepackage[normalem]{ulem} 
\usepackage{graphicx} 
\usepackage{subfigure} 
\usepackage{amssymb}  
\usepackage{amsmath}
\usepackage{mathrsfs}
\usepackage{epstopdf}
\usepackage{float}
\usepackage{booktabs} % for tabular with \toprule, \midrule, \bottomrule
\usepackage{microtype}
\makeatletter
\AtBeginDocument{%  
  \def\@makecaption#1#2{%
    \par 
    \vskip\abovecaptionskip 
    \begingroup 
     \footnotesize\rmfamily
     \sbox\@tempboxa{% 
      \let\\\heading@cr
      #1: #2%
     }%
     \@ifdim{\wd\@tempboxa >\hsize}{%
      \begingroup
       \samepage
       \flushing
       \let\footnote\@footnotemark@gobble
       #1: #2\par
      \endgroup
     }{%
       \global \@minipagefalse
       \hb@xt@\hsize{\hfil\unhbox\@tempboxa\hfil}%
     }%
    \endgroup
    \vskip\belowcaptionskip
  }%
}
\makeatother
\usepackage[colorlinks,breaklinks]{hyperref}

\newcommand{\llangle}{\langle\kern-0.2em \langle}
\newcommand{\rrangle}{\rangle\kern-0.2em \rangle}

\begin{document}
 
%%%%%%%%%%%%%%%%%%%%%%%%%%%%%%%%%%%%%%%%%%%%%%%%%%%%%%%%%%%%%%%%%%%%%%%%%%%%%%%
% Notes on Improved Generalized Harmonic Gauge Conditions for Cosmological
% Evolutions
%%%%%%%%%%%%%%%%%%%%%%%%%%%%%%%%%%%%%%%%%%%%%%%%%%%%%%%%%%%%%%%%%%%%%%%%%%%%%%%
\title{Improved Gauge Conditions for Large Expansion Cosmological
  Models}

\author{Fan Zhang\,${}^{a,b}$ and Lee Lindblom\,${}^c$}

\affiliation{${}^a$Institute for Frontiers in Astronomy and Astrophysics,
  Beijing Normal University, Beijing 102206, China}

\affiliation{${}^b$Gravitational Wave and Cosmology Laboratory,
  School of Physics and Astronomy,
  Beijing Normal University, Beijing 100875, China}

\affiliation{${}^c$ Department of Physics, University 
  of California at San Diego, La Jolla, CA 92093, USA}

\date{\today}
 
\begin{abstract}
  This study develops a new gauge condition for performing long
  duration numerical evolutions of Einstein's equation on cosmological
  spacetimes with non-trivial spatial topologies.  This new gauge
  makes it possible to perform reliable numerical evolutions on these
  manifolds in which the spatial volumes become more than one million
  times larger. This new condition was tested for this study by
  evolving homogeneous and isotropic initial data on exemplar
  manifolds with three spatial topologies: Sixth-Turn space, the lens
  space $L(5,2)$, and the Seifert fiber space
  $SFS[S2:(2,1)(2,1)(2,-1)]$.
\end{abstract}
 
\maketitle

%%%%%%%%%%%%%%%%%%%%%%%%%%%%%%%%%%%%%%%%%%%%%%%%%%%%%%%%%%%%%%%%%%%%%%%%%%%%%%%
% Introduction
%%%%%%%%%%%%%%%%%%%%%%%%%%%%%%%%%%%%%%%%%%%%%%%%%%%%%%%%%%%%%%%%%%%%%%%%%%%%%%%
\section{Introduction}
\label{s:Introduction}

Recent studies of homogeneous and isotropic cosmological models on
manifolds with non-trivial spatial topologies used a simple harmonic
gauge condition in their numerical solutions of the Einstein evolution
equations~\cite{Zhang2026a, Zhang2026b, Zhang2026c}. The spacetimes
produced by those studies could be evolved long enough for their
spatial volumes to expand by about a factor of ten.  This technical
study is focused on the development and testing of a new gauge
condition that allows reliable numerical evolutions of simple
cosmological models that achieve much larger spatial expansions on
spacetimes with non-trivial topologies.

Section~\ref{s:CosmoGaugeCondition} describes how gauge choices are
made in the generalized harmonic representation of the Einstein
evolution equations used in this study.  The properties of the gauge
source vector developed originally for binary black hole
mergers~\cite{Lindblom2009, Szilagyi:2009qz} and ultrarelativistic
particle collisions~\cite{Choptuik2010} are used to motivate a new
class of gauge conditions that allow the lapse and the cosmological
scale factor to grow at about the same rate in these cosmological
evolutions.  Keeping these quantities commensurate in size prevents
the spacetime metric from becoming ill conditioned and difficult to
invert numerically.

Section~\ref{s:SimpleInitialData} describes the initial data for the
Einstein equations used in this study to perform evolutions of simple
cosmological models that test the new gauge condition.  These
homogeneous and isotropic initial data were prepared on manifolds with
spatial slices having three different non-trivial topologies:
Sixth-Turn space (which admits a flat geometry), the lens space
$L(5,2)$, and the Seifert fiber space $ SFS[S2:(2,1)(2,1)(2,-1)]$.

Section~\ref{s:LargeExpansionCosmologies} describes the numerically
convergent evolution of the initial data constructed in
Sec.~\ref{s:SimpleInitialData} using the new gauge condition. The
spatial volumes in these evolutions became larger by a factor of a few
million before the evolutions were terminated when the constraints
started to grow unreliably.  The resulting evolutions are shown to be
numerically convergent and to be homogeneous and isotropic.

Section~\ref{s:Discussion} provides a brief summary and discussion of
the results. Suggestions are also presented for further improvements
that might allow evolutions with even larger volume expansions.

%%%%%%%%%%%%%%%%%%%%%%%%%%%%%%%%%%%%%%%%%%%%%%%%%%%%%%%%%%%%%%%%%%%%%%%%%%%%%
% Cosmological Gauge Conditions
%%%%%%%%%%%%%%%%%%%%%%%%%%%%%%%%%%%%%%%%%%%%%%%%%%%%%%%%%%%%%%%%%%%%%%%%%%%%%
\section{Cosmological Gauge Conditions}
\label{s:CosmoGaugeCondition}
%%%%%%%%%%%%%%%%%%%%%%%%%%%%%%%%%%%%%%%%%%%%%%%%%%%%%%%%%%%%%%%%%%%%%%%%%%%%%

The spacetime coordinates in the generalized harmonic representation
of Einstein's equation are determined by the choice of a gauge source
vector, $H_a$.  This representation of Einstein's equation is
symmetric hyperbolic (and therefore has a well posed initial value
problem) when $H_a$ is any function of the spacetime metric, but not
its derivatives.  The basic features of the covariant generalized
harmonic representation of Einstein's equation used in this study are
summarized in Appendix~\ref{s:SymmetricHyperbolicSystem}.

It will be useful in this discussion to represent the spacetime metric
$\psi_{ab}$ in the 3+1 ADM~\cite{ADM1962} form
\begin{eqnarray}
  ds^2 &=& \psi_{ab}\,dx^a dx^b,\nonumber \\ &=& -N^2 dt^2 +
  g_{ij}\left(dx^i + N^i dt\right)\left(dx^j + N^j dt\right),
  \label{e:ADMMetricDef}
\end{eqnarray}
where $g_{ij}$ is the spatial metric, $N$ the lapse, and $N^i$ the
shift.\footnote{In this study tensor indices from the first part of
the alphabet, $a$, $b$, $c$, $...$, represent four-dimensional 
spacetime coordinates, while tensor indices from the middle of the
alphabet, $i$, $j$, $k$, $...$, represent spatial coordinates.} In
cosmological models the spatial metric $g_{ij}$ becomes larger,
ultimately becoming very large, as the spatial volume of the universe
expands.  This can and will cause stability problems in numerical
evolutions unless the gauge makes the lapse $N$ grow at a rate
commensurate with the growth of the length scales determined by the
spatial metric $g_{ij}$.

Previous studies of homogeneous and isotropic cosmological models on
manifolds with non-trivial topologies~\cite{Zhang2026a, Zhang2026c}
used the simple harmonic gauge condition, $H_a=0$.  Those studies
showed that evolutions using harmonic gauge make the lapse grow
rapidly: $N(t)\approx a^3(t)$, where $a(t)$ is the scale factor
defined by
\begin{equation}
  a^6(t) = \frac{\det g_{\,ij}(t)}{\det g_{\,ij}(0)}.
  \label{e:ScaleFactorDef}
\end{equation}
Therefore harmonic gauge in these cosmological models makes the lapse
grow much faster than the spatial scale factor.  This mismatch leads
to numerical instabilities (e.g. the spacetime metric becomes
difficult to invert numerically) that limit the duration and the
spatial expansion that can be achieved by numerical cosmological
models using harmonic gauge.

The time component of the generalized harmonic gauge condition used in
numerical binary black-hole evolutions~\cite{Lindblom2009,
  Szilagyi:2009qz} and ultrarelativistic particle
collisions~\cite{Choptuik2010} has the form
\begin{eqnarray}
  t^aH_a &=& \mu \log\left(\frac{N(t)}{\sqrt{\det g_{\,ij}(t)}}\right),
  \label{e:BBHGauge}
\end{eqnarray} 
where $\mu$ is a constant, and $t^a$ is the future directed unit
normal to the $t=$ constant hypersurfaces.  This gauge has the effect
of limiting the growth in the ratio $N/\sqrt{\det g_{\,ij}}$, which is
essential for obtaining numerically stable evolutions that include
binary black-hole mergers.  In the cosmological setting this gauge
condition, Eq.~(\ref{e:BBHGauge}), can be re-written in the form
\begin{eqnarray}
  t^aH_a &=& \mu \log\left[\frac{N(t)}{a(t)^3}\right]
  -\mu\log\sqrt{\det g_{ij}(0)},
  \label{e:AltBBHGauge}
\end{eqnarray} 
which limits the growth of $N/a^3$. Unfortunately this results in
evolutions with $N\propto a^3$, similar to those obtained with simple
harmonic gauge.  Numerical tests conducted for this study confirm
this: Evolutions of homogeneous and isotropic cosmological models on
manifolds with non-trivial spatial topologies using the
Eq.~(\ref{e:BBHGauge}) gauge were not successful in extending the
spatial expansions achieved with harmonic gauge.

This study proposes a new gauge condition for cosmological spacetime
evolutions that generalizes the condition designed for binary black-hole
evolutions in Eq.~(\ref{e:AltBBHGauge}).  This new condition is given by
\begin{eqnarray}
  t^aH_a &=&
 \mu \log\left[\frac{N(t)}{a^\sigma(t)}\right],
 \label{e:NewGaugeDef}\\
 H_k &=& 0,
 \label{e:NewHkGaugeDef}
\end{eqnarray} 
where $\mu$ and $\sigma$ are constants.  This condition is designed to
drive the evolution of the lapse $N$ toward keeping the ratio
$N(t)/a^\sigma(t)$ fixed on a timescale set by the damping parameter
$\mu$.  The $\sigma=3$ version of this condition is roughly equivalent
to Eq.~(\ref{e:AltBBHGauge}).  The case $\sigma=1$ is the one expected
to be most useful for the numerical evolutions of homogeneous and
isotropic cosmological models where it is important for $N(t)$ and
$a(t)$ to maintain comparable sizes.

\begin{figure}[!h]
{\centering
\includegraphics[height=0.32\textwidth]{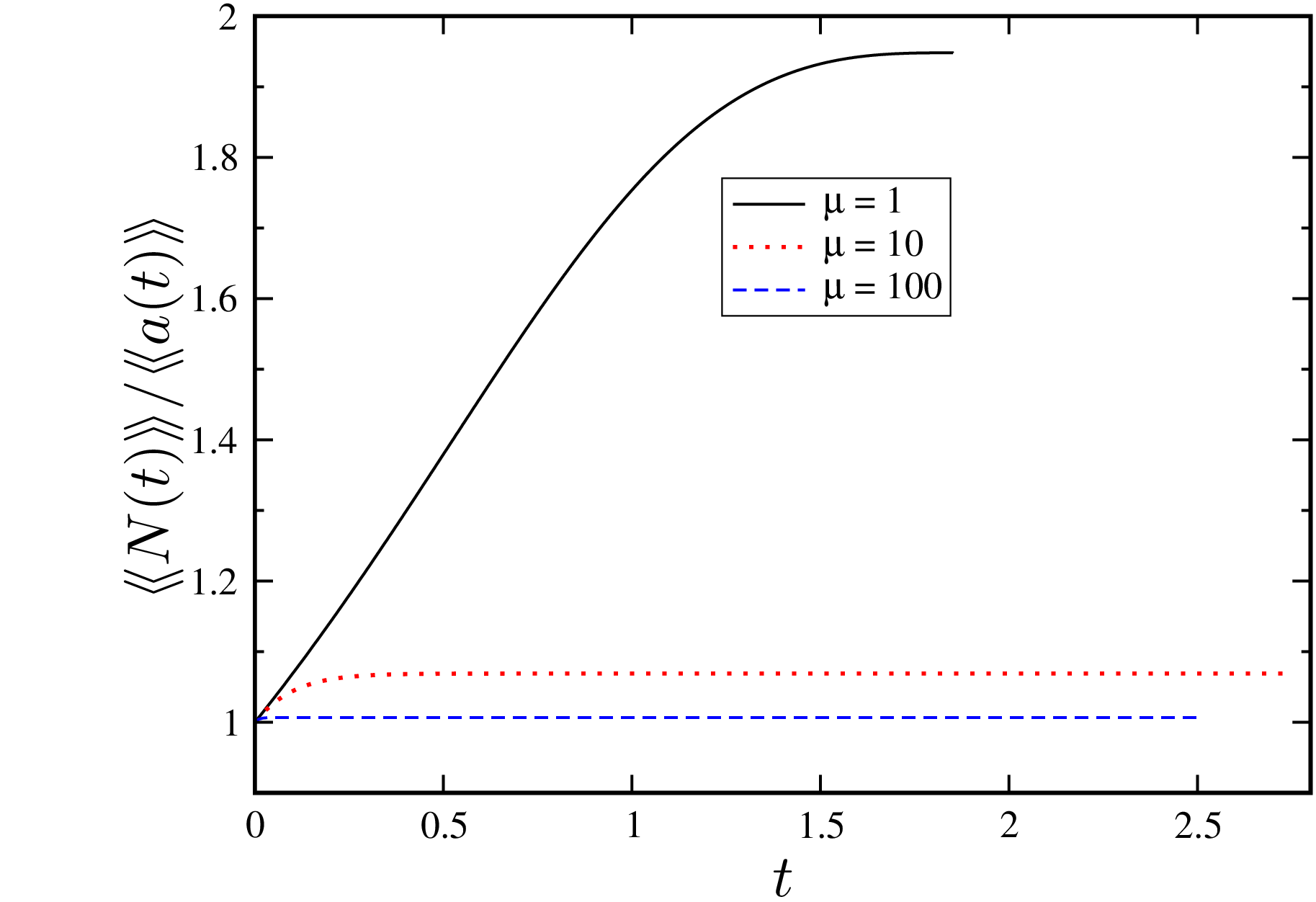}}
\caption{\label{f:QScaleRatio} This figure illustrates the ratios
  of the spatial average of the lapse $\llangle
  N(t)\rrangle$ to the spatial average of the scale factor
  $\llangle a(t) \rrangle$ for homogeneous and isotropic
  evolutions on Sixth-Turn space using the gauge condition defined
  in Eq.~(\ref{e:NewGaugeDef}) with $\sigma=1$ and various values of
the damping parameter $\mu$.}
\end{figure}

Numerical tests of the new gauge condition, Eqs.~(\ref{e:NewGaugeDef})
and (\ref{e:NewHkGaugeDef}),
with $\sigma=1$ were conducted for this study to explore the effects
of different values of the damping parameter $\mu$.  Those tests were
performed (using the methods described in detail in
Secs.~\ref{s:SimpleInitialData} and \ref{s:LargeExpansionCosmologies})
by evolving homogeneous and isotropic initial data on Sixth-Turn
space.  The evolutions in these tests were continued until the
Einstein constraints started to become large.
Figure~\ref{f:QScaleRatio} shows that the ratio $\llangle
N(t)\rrangle/\llangle a(t)\rrangle$ is driven in these test evolutions
toward order unity constant values, where $\llangle N(t)\rrangle$ and
$\llangle a(t)\rrangle$ are the spatial averages defined by
\begin{eqnarray}
  \llangle N(t)\,\rrangle
  &=& \frac{1}{\mathcal{V}(t)}\int N(t)\,\sqrt{g}\,d^{\,3}x,
  \label{e:LapseAve}\\
  \llangle a(t)\,\rrangle
  &=& \frac{1}{\mathcal{V}(t)}\int a(t)\,\sqrt{g}\,d^{\,3}x,
  \label{e:ScaleFactorAve}
\end{eqnarray} 
and $\mathcal{V}$ is the spatial volume of the $t=$ constant
hypersurfaces defined by
\begin{eqnarray}
  \mathcal{V}(t) &=& \int \sqrt{g}\,d^{\,3}x.
  \label{e:Volume}
\end{eqnarray}
This ratio becomes closer to unity as the value of $\mu$ increases.

\begin{figure}[!h]
{\centering
\includegraphics[height=0.32\textwidth]{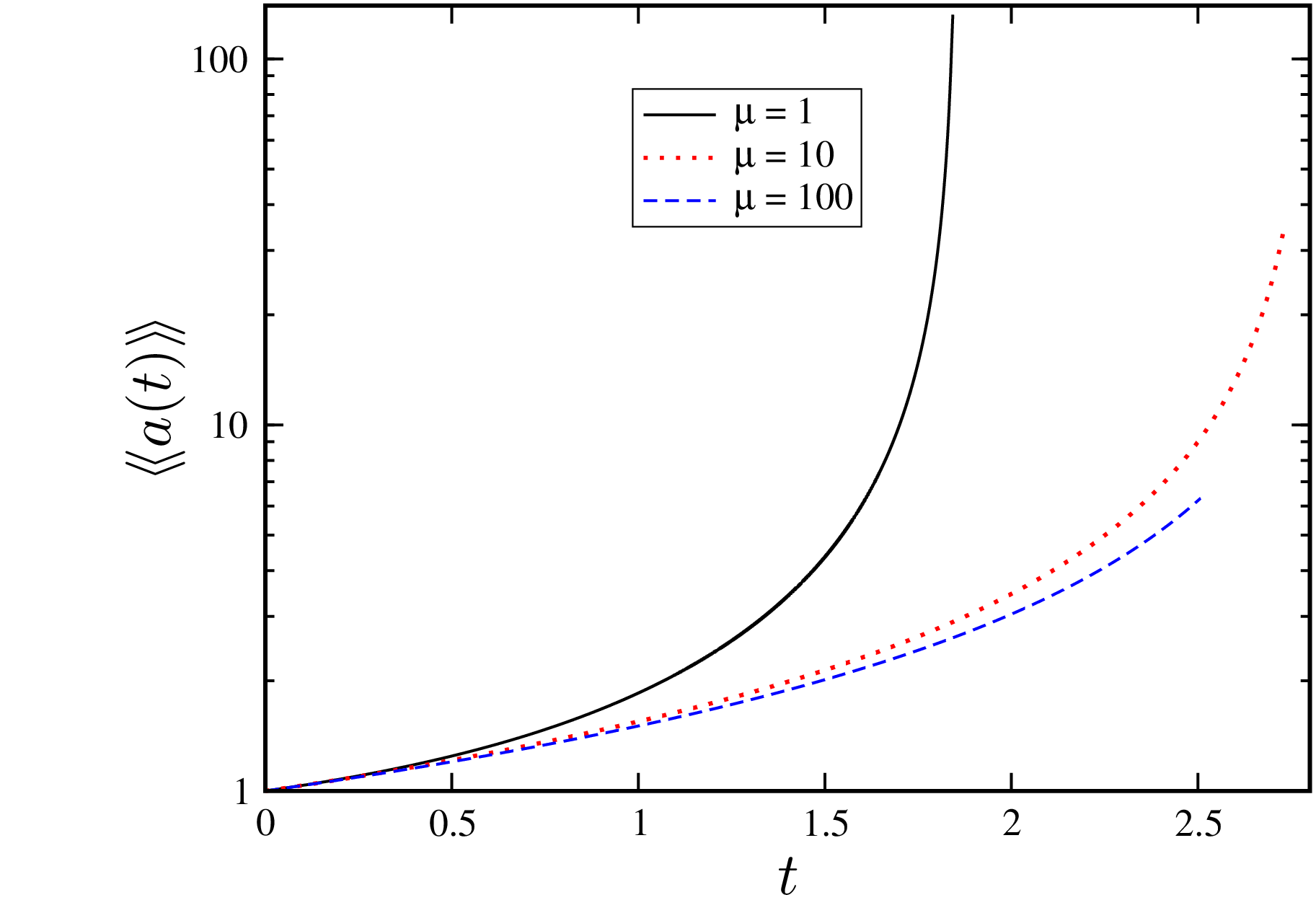}}
\caption{\label{f:QScaleAve} This figure illustrates the spatial
  average of the scale factor $\llangle a(t) \rrangle$ for homogeneous
  and isotropic evolutions on Sixth-Turn space using the gauge
  condition defined in Eqs.~(\ref{e:NewGaugeDef}) and
  (\ref{e:NewHkGaugeDef}) with $\sigma=1$ and various values of the
  damping parameter $\mu$.}
\end{figure}

Figure~\ref{f:QScaleAve} shows that the spatial average of the scale
factor, $\llangle a(t)\rrangle$, grew to the largest value with
$\max\llangle a(t)\rrangle\approx 132$ for the test with $\mu=1$
before the evolutions started to become unreliable.  The goal of this
study is to develop methods for constructing long duration
cosmological evolutions numerically, so the $\mu=1$ parameter choice
is optimal for this purpose.
%

%%%%%%%%%%%%%%%%%%%%%%%%%%%%%%%%%%%%%%%%%%%%%%%%%%%%%%%%%%%%%%%%%%%%%%%%%%%%%
% Simple Cosmological Initial Data
%%%%%%%%%%%%%%%%%%%%%%%%%%%%%%%%%%%%%%%%%%%%%%%%%%%%%%%%%%%%%%%%%%%%%%%%%%%%%
\section{Simple Cosmological Initial Data}
\label{s:SimpleInitialData}
%%%%%%%%%%%%%%%%%%%%%%%%%%%%%%%%%%%%%%%%%%%%%%%%%%%%%%%%%%%%%%%%%%%%%%%%%%%%%

The new generalized harmonic gauge conditions described in
Sec.~\ref{s:CosmoGaugeCondition} were tested for this study by
evolving simple homogeneous cosmological initial data on three
manifolds having non-trivial spatial topologies: Sixth-Turn space,
$L(5,2)$, and $SFS[S2:(2,1)(2,1)(2,-1)]$. Sixth-Turn space (also
called $E5$) is one of the compact orientable three-dimensional
manifolds that admits a flat geometry.  $L(5,2)$ is one of the lens
spaces constructed by identifying points on the three-sphere that are
related by a discrete isometry.  $SFS[S2:(2,1)(2,1)(2,-1)]$ is a
Seifert fiber space constructed over the two-sphere.  Einstein's
equation is solved numerically on these manifolds using the multi-cube
structures (i.e. collections of non-overlapping cubic coordinate
patches) for these spaces constructed in Ref.~\cite{Lindblom2022}.

The differentiable structures that determine the continuity of vectors
and tensors across the interfaces between the coordinate patches are
defined using the $\mathcal{C}^{2-}$ reference metrics, $\tilde
g_{ij}$, constructed numerically for these manifolds in
Ref.~\cite{Lindblom2022}.  These reference metrics were designed to be
sufficiently smooth at the interfaces between cubic patches to define
the differentiable structures.  But they were not designed to be
uniform geometrically or to solve the Einstein constraint
equations. Consequently they do not make suitable initial data for
cosmological models.

Spatially uniform geometries can be constructed from the reference
metrics, $\tilde g_{ij}$ by solving the Ricci flow equations on these
manifolds, see Ref.~\cite{Zhang2026c}.  The resulting homogeneous
geometries can then be used as conformal metrics in the construction
of simple solutions to the constant-mean-curvature form of the
Einstein constraint equations, see Ref.~\cite{Zhang2022} for
details. The resulting solution provides a physical spatial metric,
$g_{ij}$, and an extrinsic curvature trace, $K$, that satisfy the
Einstein constraints on a $t=$ constant hypersurface in a spacetime
with a cosmological constant but no other matter fields.  The value
of the extrinsic curvature trace, $K$, can be chosen freely in
these simple solutions, but the cosmological constant, $\Lambda$, is
fixed by the equation,
\begin{equation}
  \Lambda = \textstyle\frac{1}{3}K^2+\textstyle\frac{1}{2}\llangle R\, \rrangle,
  \label{e:CMCParameterChoices}
\end{equation}
where $\llangle R\, \rrangle$ is the spatial average of the Ricci
scalar curvature $R$ associated with $g_{ij}$,
\begin{eqnarray}
  \llangle R\,\rrangle
  &=& \frac{1}{\mathcal{V}}\int R\, \sqrt{g}\,d^{\,3}x.
  \label{e:RicciScalarAve}
\end{eqnarray}

\begin{figure}[!h]
{\centering
\includegraphics[height=0.32\textwidth]{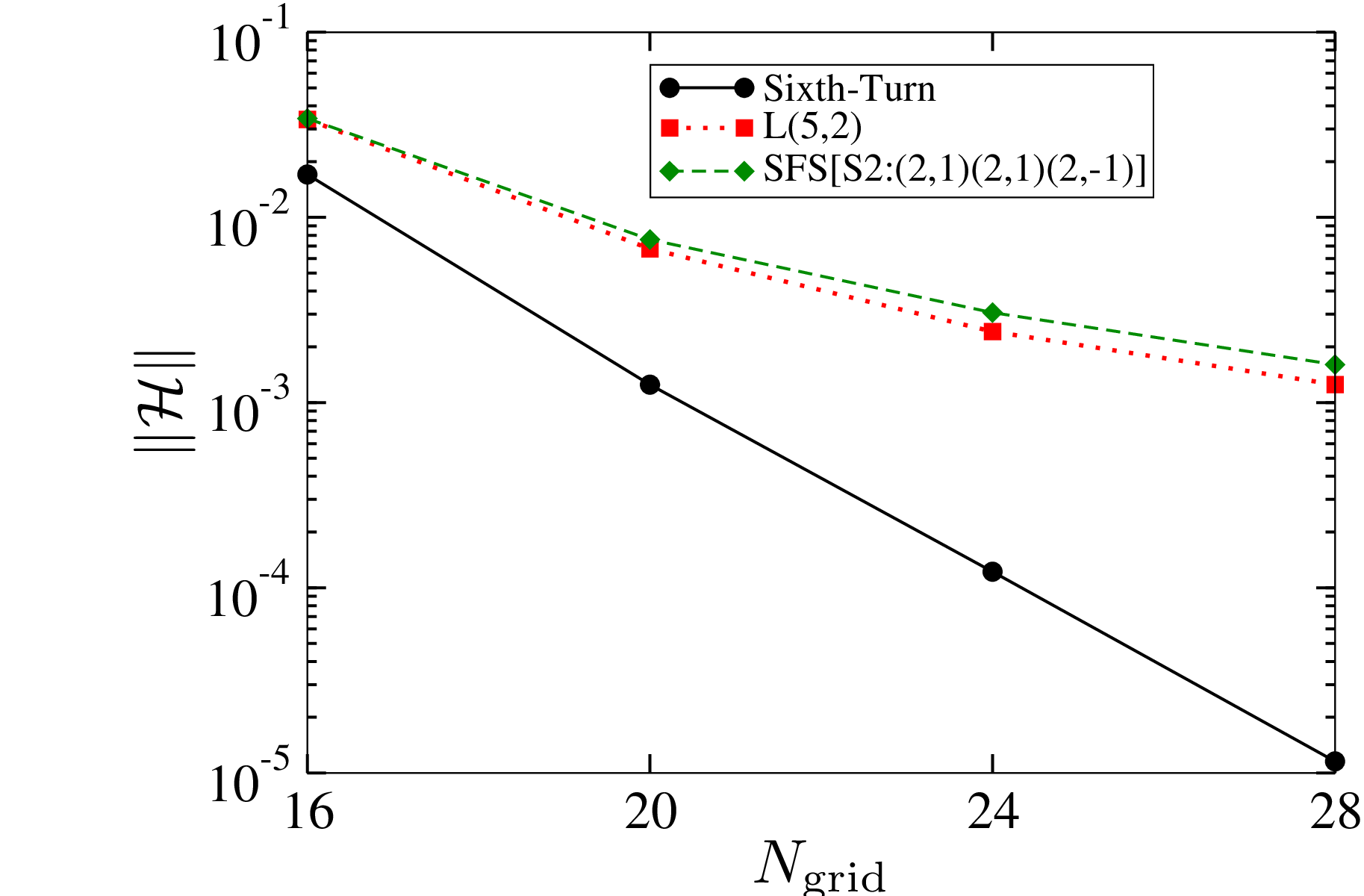}}
\caption{\label{f:HamilgonianConvergence} This figure illustrates the
  the norm $\|\mathcal{H}\|$ of the Hamiltonian constraint defined in
  Eq.~(\ref{e:EinsteinConstraint}) as a function of the numerical grid
  size, $N_\mathrm{grid}$, for solutions of the Einstein initial value
  constraint equations on the Sixth-Turn, $L(5,2)$, and
  $SFS[S2:(2,1)(2,1)(2,-1)]$ manifolds.}
\end{figure}

Numerical solutions to the Ricci flow equation, and the Einstein
conformal initial value equations were constructed for this study
using the \verb!SpEC! numerical relativity code (developed originally
by the Caltech and Cornell numerical relativity groups). Homogeneous
and isotropic initial data for Einstein's equation were constructed in
Ref.~\cite{Zhang2026c} for Sixth-Turn space and $L(5,2)$.  The same
numerical methods have been used in this study to construct initial
data for $SFS[S2:(2,1)(2,1)(2,-1)]$.  These numerical solutions were
constructed on pseudo-spectral grids in each multi-cube coordinate
patch, which have $N_\mathrm{grid}$ collocation points in each
direction in each patch.  The accuracy of the numerical solutions was
monitored by evaluating the norm of the Hamiltonian initial value
constraint, $\|\mathcal{H}\|$.  For these simple cosmological initial
data the expression for $\|\,\mathcal{H}\,\|$ can be written as
\begin{equation}
  \|\,\mathcal{H}\,\|^2 = \mathcal{V}^{-1}\int \bigl(R -
  \llangle R\, \rrangle\bigr)^2 \!\sqrt{g}\, d^{\,3}x,
  \label{e:EinsteinConstraint}
\end{equation}
see Refs.\cite{Zhang2022, Zhang2026a}.
Figure~\ref{f:HamilgonianConvergence} illustrates the convergence of
$\|\mathcal{H}\|$ as the resolution, $N_\mathrm{grid}$, is increased.
These results show that the initial data computed for this study are
numerically convergent and reasonably accurate.  The physical
properties, $\llangle R\,\rrangle$, $\mathcal{V}$ and $\Lambda$, of
the numerical solutions with $K=-1$ are given in Table~\ref{t:TableI}.
\begin{table}[!hbt]
  \caption{Physical properties of the initial data for the manifolds
    included in this study: the average scalar curvature $\llangle
    R\,\rrangle$ defined in Eq.~(\ref{e:RicciScalarAve}), the physical
    volume $\mathcal{V}$ defined in Eq.~(\ref{e:Volume}), and the
    cosmological constant $\Lambda$ determined by
    Eq.~(\ref{e:CMCParameterChoices}) for the geometries with $K=-1$.
    \label{t:TableI} }
  \begin{center}
  \begin{tabular}{lrlrr}
    Manifold && $\!\!\!\!\!\!
    \llangle R\, \rrangle$ \!\!\!\!\!\!
    & $\mathcal{V}$\,\,\,\,\,\,\,  & $\Lambda$\,\,\,\,\,\,\,
 \\
 \hline
 \vspace{-7pt}\\
 Sixth-Turn
 &   \!\! -1.&\!\!\!7538$\times 10^{-8}$\!\!\!\!\! & \!\! 6.2804 &  0.3333 \\
 $L(5,2)$
 &   \!\! 3.&\!\!\!9873 & \, 7.2874   &   2.3270 \\
 $SFS[S2:(2,1)(2,1)(2,-1)]$
 &   \!\! 2.&\!\!\!2410 & \, 10.8091  &  1.4538 \\
 \hline
  \end{tabular}
  \end{center} 
\end{table}
%

%%%%%%%%%%%%%%%%%%%%%%%%%%%%%%%%%%%%%%%%%%%%%%%%%%%%%%%%%%%%%%%%%%%%%%%%%%%%%
% Large Expansion Cosmological Models
%%%%%%%%%%%%%%%%%%%%%%%%%%%%%%%%%%%%%%%%%%%%%%%%%%%%%%%%%%%%%%%%%%%%%%%%%%%%%
\section{Large Expansion Cosmological Models}
\label{s:LargeExpansionCosmologies}
%%%%%%%%%%%%%%%%%%%%%%%%%%%%%%%%%%%%%%%%%%%%%%%%%%%%%%%%%%%%%%%%%%%%%%%%%%%%%

The representation of Einstein's equation used to perform the
numerical evolutions in this study is the covariant first-order
symmetric-hyperbolic representation developed in
Ref.~\cite{Lindblom2014}.  The dynamical fields in this representation
are the spacetime metric, $\psi_{ab}$, its time derivative,
$\Pi_{ab}=-t^c\tilde\nabla_c\psi_{ab}$, and its spatial derivatives
$\Phi_{iab}=\tilde\nabla_i\psi_{ab}$, where $t^c$ is the future
directed timelike unit normal ($t^a t^b \psi_{ab}=-1$) to the $t =$
constant hypersurfaces.  The derivative, $\tilde\nabla_a$, used in
these definitions is the torsion-free covariant derivative compatible
with the reference metric $\tilde \psi_{ab}$ that defines the
differentiable structure in each manifold:
$\tilde\nabla_c\tilde\psi_{ab}=0$. The important features of this
representation of Einstein's equation (including expressions for all
of the constraints) is summarized more completely in
Appendix~\ref{s:SymmetricHyperbolicSystem}.

The reference metric $\tilde \psi_{ab}$
used for the evolutions in this study is given by
\begin{equation}
  d\tilde s^2 = -dt^2 + g^0_{ij} dx^i dx^j,
  \label{e:RefMetricDef}
\end{equation}
where $g^0_{ij}$ is the spatial metric of the simple homogeneous and
isotropic cosmological models constructed by solving the Einstein
initial value problem in Sec.~\ref{s:SimpleInitialData}.  The values
of the dynamical fields on the initial spacelike hypersurface used to
start the evolutions are given by,
\begin{eqnarray}
  \psi^0_{ab}&=& \tilde\psi_{ab},
  \label{e:Initialpsi}\\
  \Pi^0_{tt}&=& -2K^0,\label{e:InitialPitt2}\\
  \Pi^0_{tk}&=& 0.\label{e:InitialPitk2}\\
  \Pi^0_{ij}&=&\tfrac{2}{3}K^0 g^0_{ij},
  \label{e:InitialPiij}\\
  \Phi^0_{iab}&=&0,
  \label{e:InitialPhi}
\end{eqnarray}
where $K^0$ is the trace of the extrinsic curvature used to compute
the initial data.  Appendix~\ref{s:InitialGauge} gives a description
of how the choice of gauge source vector $H_a$ effects the values of
$\Pi^0_{ab}$ on the initial hypersurface.\footnote{Note that the
expressions given in Ref.~\cite{Zhang2026a} for the initial values of
$\Pi^0_{ab}$ contained typographical errors.  The correct expressions
are given in Eqs.~(\ref{e:InitialPitt2})--(\ref{e:InitialPiij}) of
this paper.  Fortunately the code used to construct the data reported
in Ref.~\cite{Zhang2026a} used the correct expressions, so the results
reported there were not effected by this error.}

Evolutions using the new generalized harmonic gauge conditions
described in Sec.~\ref{s:CosmoGaugeCondition} and the simple
cosmological initial data described in Sec.~\ref{s:SimpleInitialData}
were carried out in this study on three manifolds having non-trivial
spatial topologies: Sixth-Turn space, $L(5,2)$, and
$SFS[S2:(2,1)(2,1)(2,-1)]$. These evolutions were carried out using
the \verb!SpEC! numerical relativity code which uses pseudo-spectral
methods to evaluate spatial partial derivatives and integrals.
Adaptive time stepping for the evolutions in this study was performed
using the eighth order Dormand-Prince algorithm~\cite{Dormand1987,
  Dormand1996}.

\begin{figure}[!h]
{\centering
\includegraphics[height=0.32\textwidth]{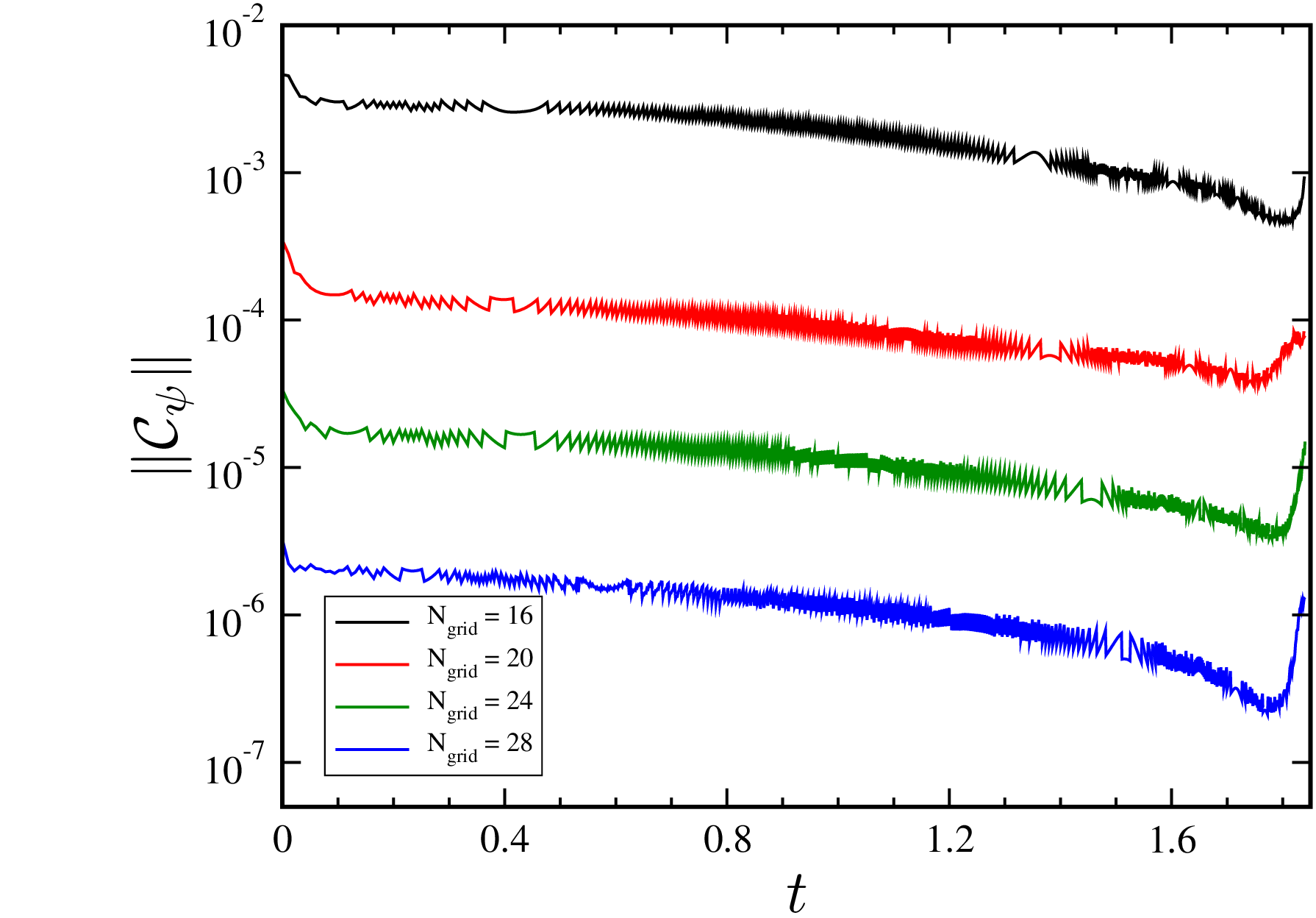}}
\caption{\label{f:E5_GhCeCov} This figure illustrates the
  dimensionless constraint norm $\|\mathcal{C}_\psi\|$
  defined in Eq.~(\ref{e:CpsiNormDef}) for evolutions on the
  Sixth-Turn (E5) manifold.}
\end{figure}

\begin{figure}[!h]
{\centering
\includegraphics[height=0.32\textwidth]{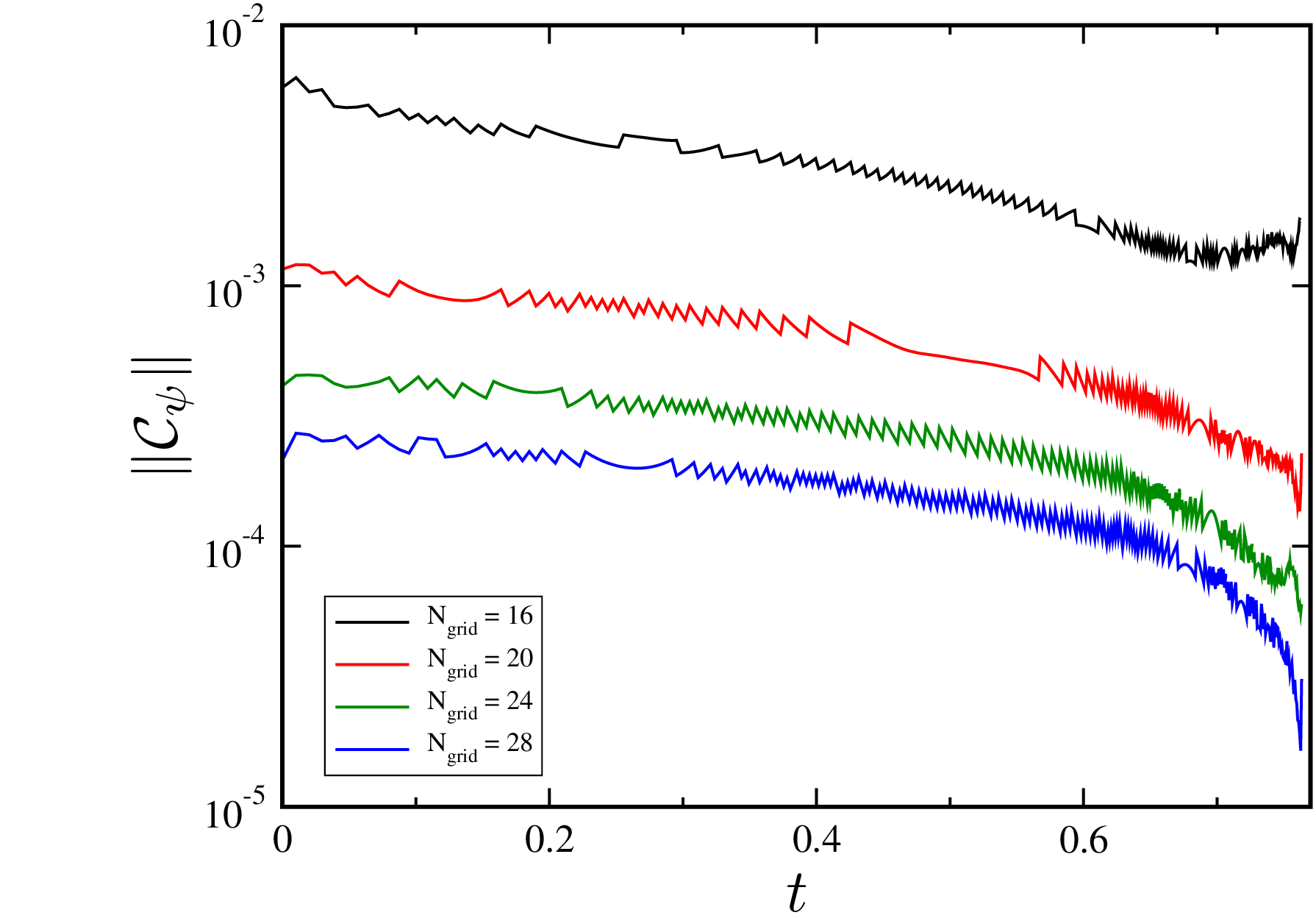}}
\caption{\label{f:L52_GhCeCov} This figure illustrates the dimensionless
  constraint norm $\|\mathcal{C}_\psi\|$ defined in Eq.~(\ref{e:CpsiNormDef})
for evolutions on the $L(5,2)$ manifold.}
\end{figure}

\begin{figure}[!h]
{\centering
\includegraphics[height=0.32\textwidth]{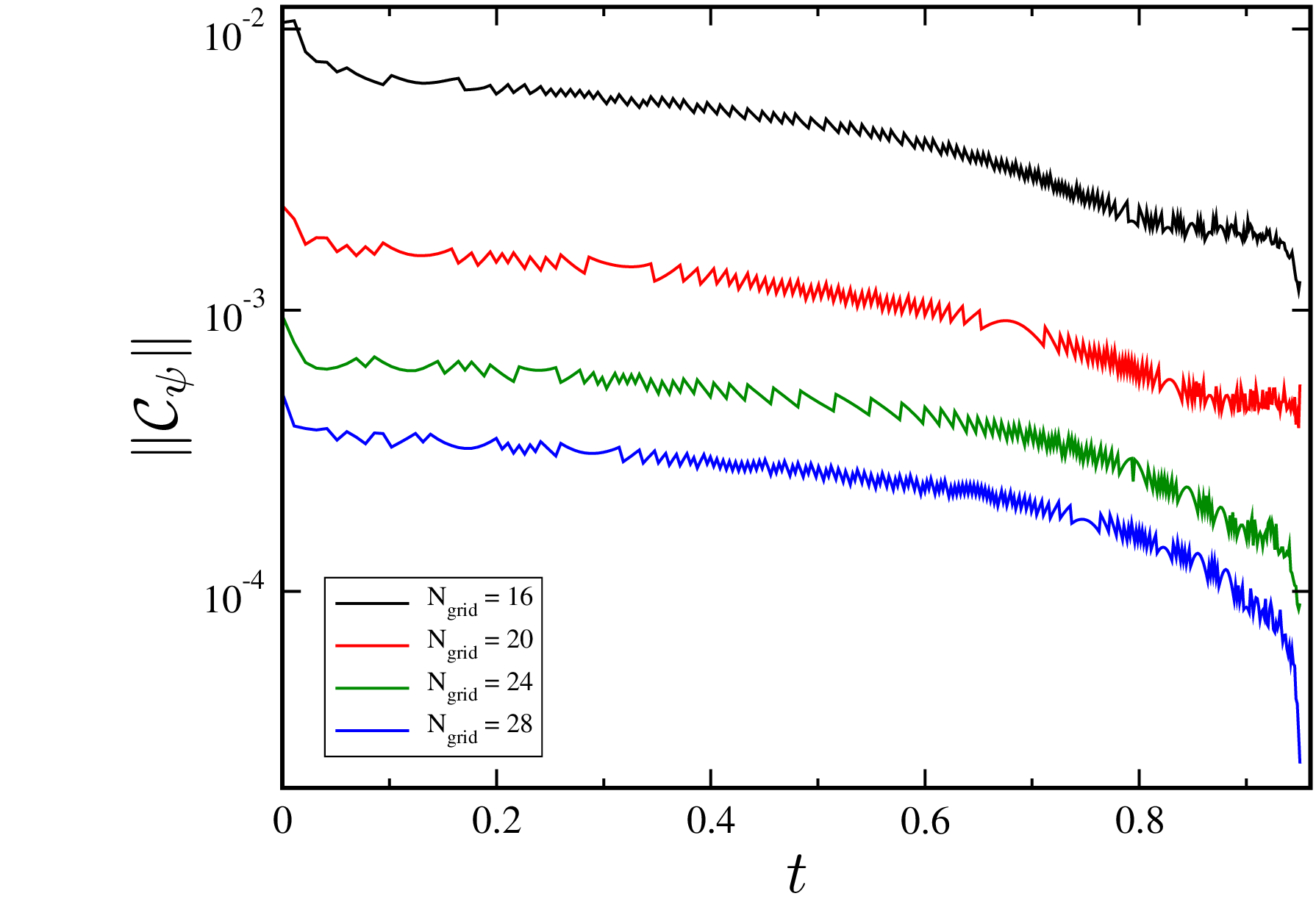}}
\caption{\label{f:SFS_GhCeCov} This figure illustrates the dimensionless
  constraint norm $\|\mathcal{C}_\psi\|$ defined in Eq.~(\ref{e:CpsiNormDef})
for evolutions on the SFS[S2:(2,1)(2,1)(2,-1)] manifold.}
\end{figure}

The accuracy of the numerical evolutions is monitored by evaluating
the norm $\|\mathcal{C}_\psi\|$ defined in Eq.~(\ref{e:CpsiNormDef})
that measures the average fractional violations of the large number of
constraints in the first-order generalized harmonic evolution system.
Figures~\ref{f:E5_GhCeCov}--\ref{f:SFS_GhCeCov} show
$\|\mathcal{C}_\psi(t)\|$ for evolutions performed at numerical
resolutions $N_\mathrm{grid}=\{16,20,24,28\}$ on each of the manifolds
included in this study.  These results show that the numerical methods
used in this study are reasonably accurate and numerically convergent.

\begin{figure}[!b]
{\centering
\includegraphics[height=0.32\textwidth]{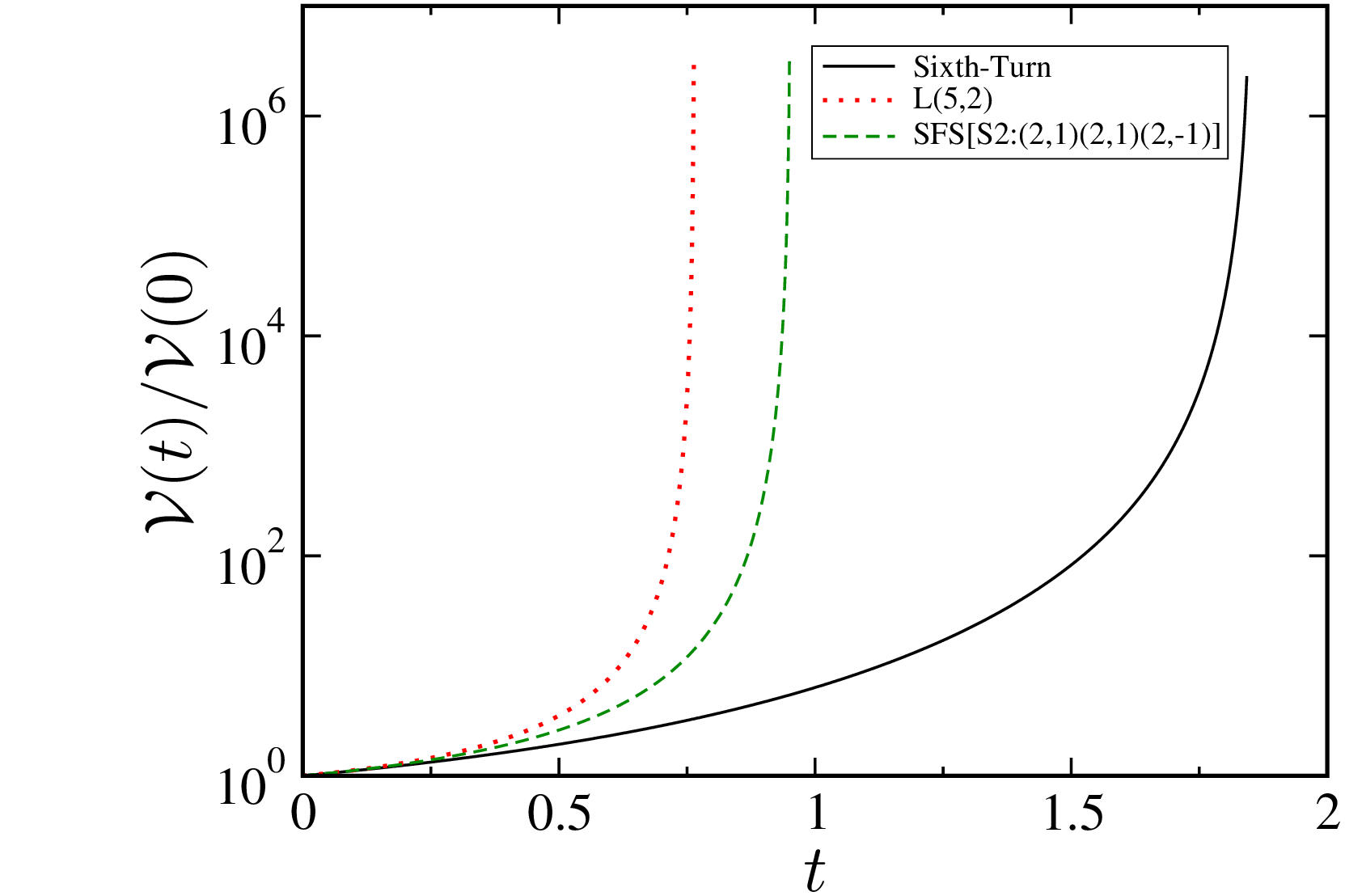}}
\caption{\label{f:Volumes} This figure illustrates the ratio of the
  spatial volume $\mathcal{V}(t)$ defined in Eq.~(\ref{e:Volume}) to
  its initial value $\mathcal{V}(0)$ for the long duration evolutions
  on the Sixth-Turn, $L(5,2)$, and the SFS[S2:(2,1)(2,1)(2,-1)]
  manifolds. }
\end{figure}

The primary motivation for this study was the development of a new
gauge condition that allows reliable evolutions of cosmological models
that can achieve large spatial expansions.  To test how well the new
gauge condition in Eqs.~(\ref{e:NewGaugeDef}) and
(\ref{e:NewHkGaugeDef}) has succeeded, the evolutions of these simple
cosmological models were continued numerically until the constraints
shown in Figs.~\ref{f:E5_GhCeCov}--\ref{f:SFS_GhCeCov} began to rise
sharply.  Figure~\ref{f:Volumes} illustrates the ratio of the spatial
volume $\mathcal{V}(t)$ defined in Eq.~(\ref{e:Volume}) to its initial
value $\mathcal{V}(0)$ for the $N_\mathrm{grid}=28$ evolutions.  (The
lower resolution results are indistinguishable from the
$N_\mathrm{grid}=28$ results on the scale of this figure.) These
results show that the spatial volumes in these tests increase by a
few million times the initial volume.

Table~\ref{t:TableII} lists the maximum value of the ratio
$\mathcal{V}(t)/\mathcal{V}(0)$ as well as the maximum value of the
spatial average of the scale factor, $\llangle a(t) \rrangle$, defined
by Eq.~(\ref{e:ScaleFactorAve}) for the $N_\mathrm{grid}=28$ evolution
on each of the manifolds.  The scale factors in these evolutions reach
values of 132 and above, which is a significant improvement over the
cosmological models computed in Refs.\cite{Zhang2026a, Zhang2026c} in
which the maximum scale factors were only about 2.15.  These results
show that the new gauge condition allows stable accurate
large-expansion cosmological evolutions on a variety of manifolds with
non-trivial topologies.
\begin{table}[!hbt]
  \caption{The maximum values of the spatial volume,
    $\max\mathcal{V}(t)/\mathcal{V}(0)$, and the maximum values of the
    scale factor, $\max\llangle a(t)\rrangle$, reached in the
    $N_\mathrm{grid}=28$ evolution on each manifold.
    \label{t:TableII} }
  \begin{center}
  \begin{tabular}{lll}
    Manifold &\, $\max\mathcal{V}(t)/\mathcal{V}(0)$ 
    & \,$\max\llangle a(t)\rrangle$ 
 \\
 \hline 
 \vspace{-7pt}\\
 Sixth-Turn
 & \quad   $2.305\times 10^{6}$  & \quad 132.10 \\
 $L(5,2)$
 & \quad   $3.152\times 10^6$  & \quad 146.63  \\
 $SFS[S2:(2,1)(2,1)(2,-1)]$
 &  \quad  $3.537\times 10^6$  & \quad 152.36 \\
 \hline
  \end{tabular}
  \end{center} 
\end{table}

\begin{figure}[!h]
{\centering
\includegraphics[height=0.32\textwidth]{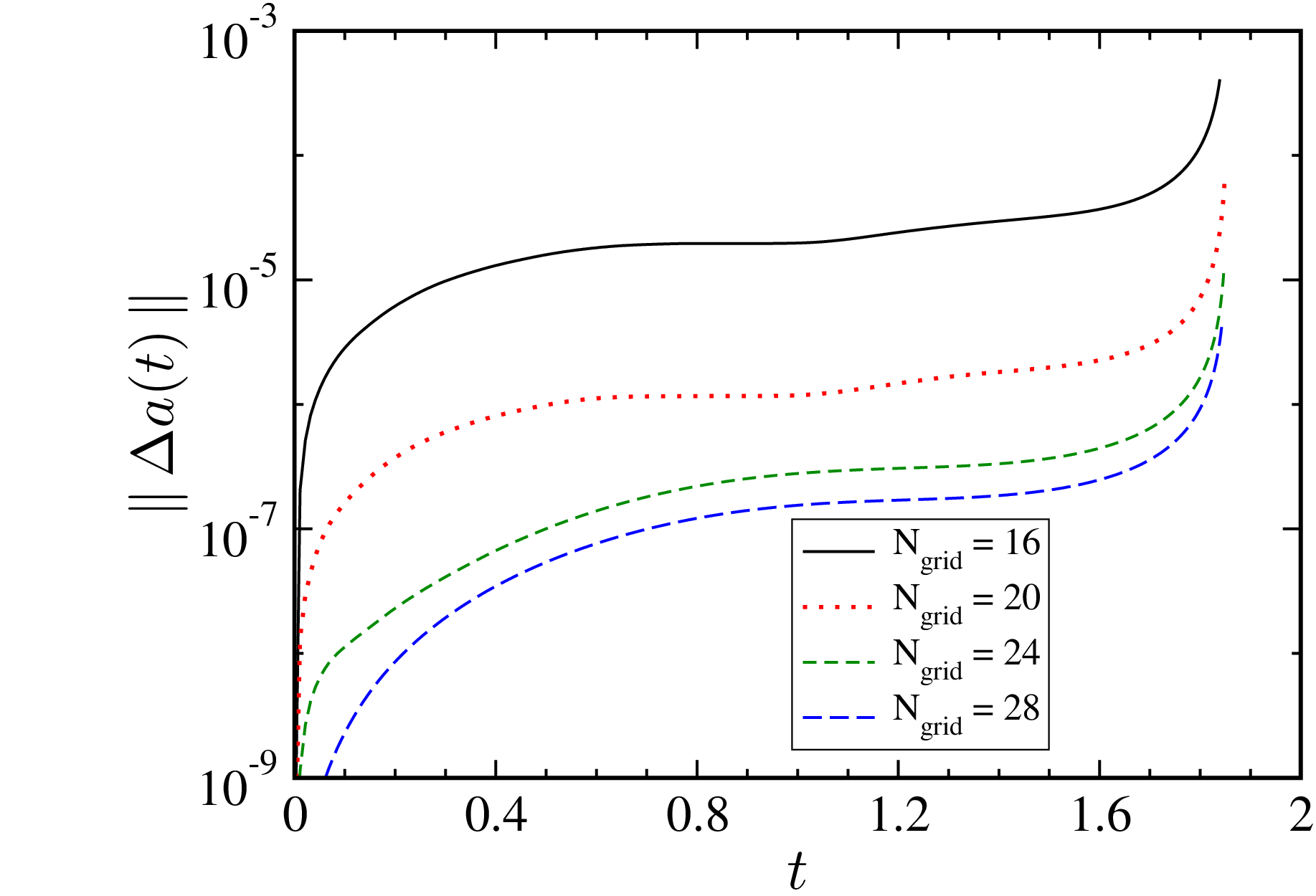}}
\caption{\label{f:E5_DeltaScaleFactor} This figure illustrates the
  variations in the scale factor $\|\Delta
  a(t)\|$ defined in Eq.~(\ref{e:DeltaScaleFactor}) for
  evolutions on the Sixth-Turn manifold.}
\end{figure}

\begin{figure}[!h]
{\centering
\includegraphics[height=0.32\textwidth]{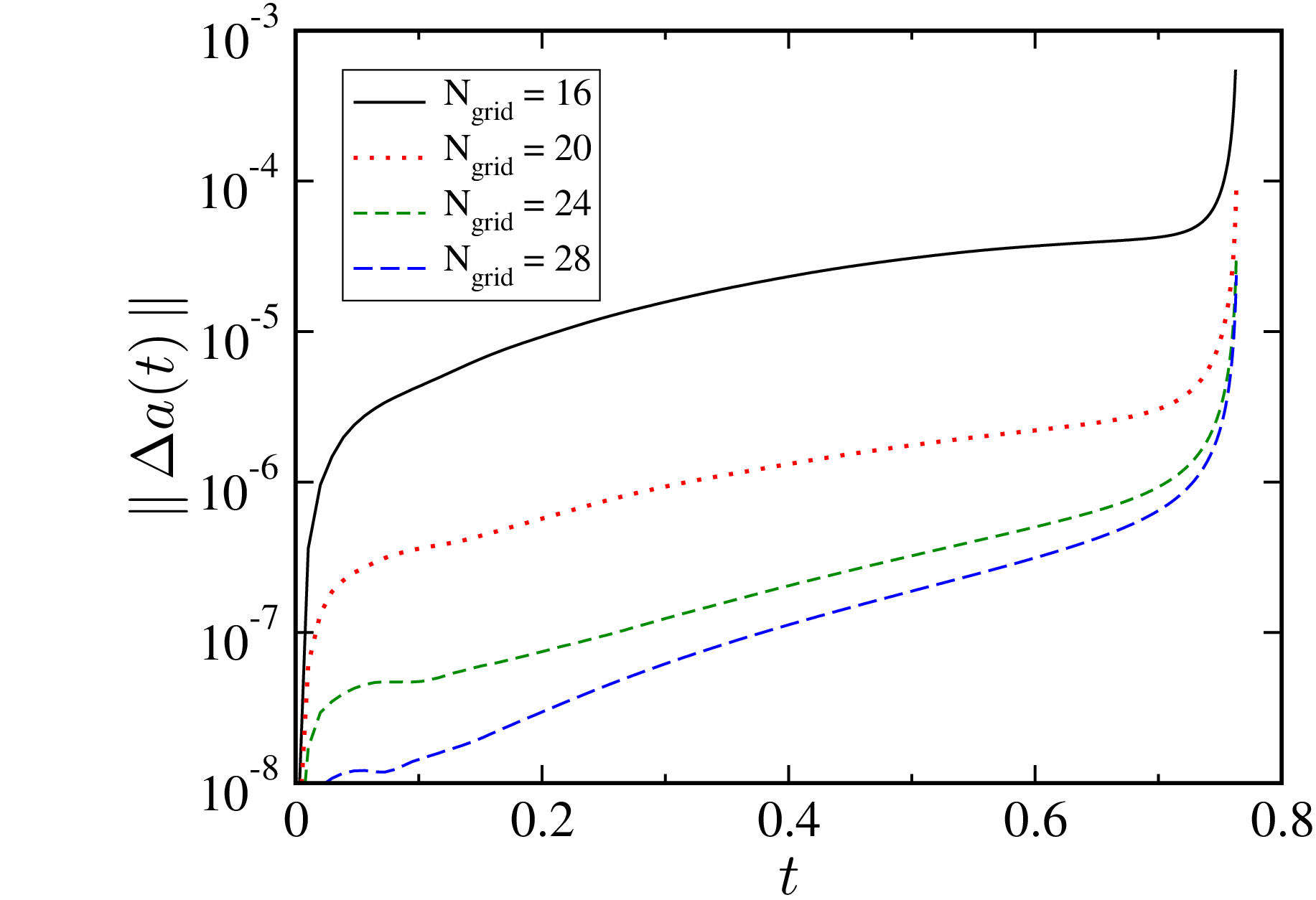}}
\caption{\label{f:L52_DeltaScaleFactor} This figure illustrates the
  variations in the scale factor $\|\Delta a(t)\|$ defined
  in Eq.~(\ref{e:DeltaScaleFactor}) for evolutions on the $L(5,2)$
  manifold.}
\end{figure}

\begin{figure}[!h]
{\centering
\includegraphics[height=0.32\textwidth]{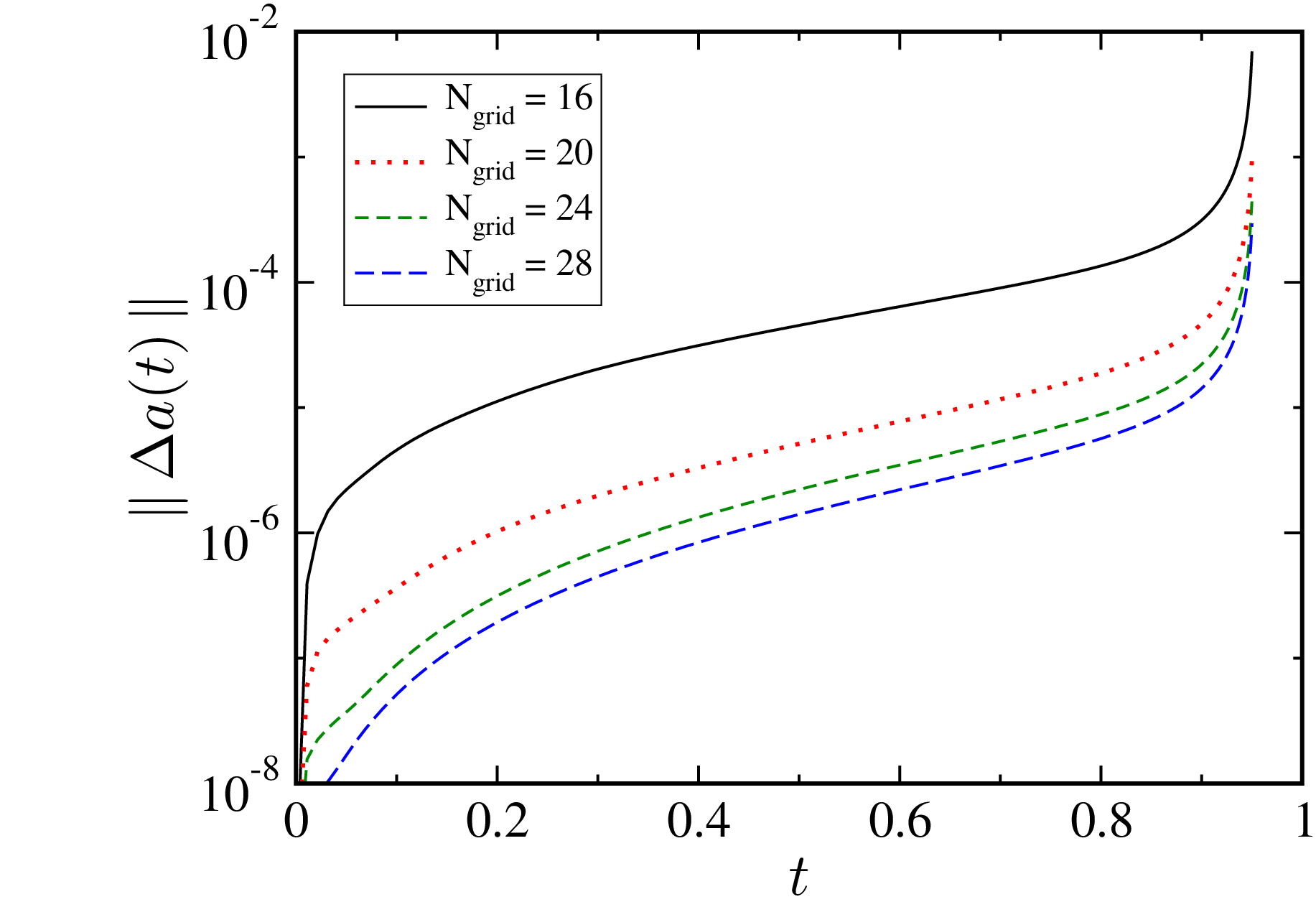}}
\caption{\label{f:SFS_DeltaScaleFactor} This figure illustrates the
  variations in the scale factor $\|\Delta a(t)\|$ defined
  in Eq.~(\ref{e:DeltaScaleFactor}) for evolutions on the
  SFS[S2:(2,1)(2,1)(2,-1)] manifold.}
\end{figure}

Several norms have been introduced to monitor the homogeneity and
isotropy of the numerical cosmological models computed for this
study.  The spatial variations in the scale factor $a(t)$, the lapse
$N(t)$, the spatial Ricci scalar curvature $R(t)$, and the curvature
scalar $\mathscr{R}(t)$ defined by
\begin{equation}
  \mathscr{R}^2 = 3 R_{ij}R^{ij}.
  \label{e:CurvatureScalarDef}
\end{equation}
have all been monitored using norms analogous to
\begin{eqnarray}
\|\,\Delta a\,\|^2 &=& 
\frac{1}{\mathcal{V}\,\llangle a \rrangle}
\int \bigl(a - \llangle a \rrangle\bigr)^2 \sqrt{g}\, d^{\,3}x.
     \label{e:DeltaScaleFactor}
\end{eqnarray}
The resulting values of these norms show that the spatial variations
in each of these quantities is small and becomes smaller as the
numerical resolution, $N_\mathrm{grid}$, of the evolution
increases. As a typical example of these results,
Figs.~\ref{f:E5_DeltaScaleFactor}--\ref{f:SFS_DeltaScaleFactor}
illustrate the results for the norm that measures the spatial
variations in the scale factor, $\|\,\Delta a(t)\|$ for the evolutions
on each of the manifolds.

The isotropy of the resulting cosmological models can be evaluated by
studying the properties of the spatial Ricci tensors on the $t=$
constant hypersurfaces.  If the Ricci tensor is determined completely
by the Ricci scalar curvature, $R_{ij}=\tfrac{1}{3}g_{ij}R$, then its
eigenvalues are identical and the geometry is isotropic.  In this case
the curvature scalar $\mathscr{R}$ defined in
Eq.~(\ref{e:CurvatureScalarDef}) equals the Ricci scalar:
$\mathscr{R}^{\,2}=R^2$.  Conversely if $\mathscr{R}^{\,2}=R^2$ then
the spatial geometry must be isotropic: Consider a frame in which the
Ricci tensor is diagonal, with eigenvalues $R_1$, $R_2$ and $R_3$.
The Ricci scalar $R$ is given in this case by $R=R_1+R_2+R_3$, while
the curvature scalar $\mathscr{R}$ is given by
$\mathscr{R}^{\,2}=3(R_1^2+R_2^2+R_3^2)$.  Using these expressions it
follows that
\begin{equation}
  \mathscr{R}^{\,2}-R^2=(R_1-R_2)^2 + (R_1-R_3)^2 + (R_2-R_3)^2.
\end{equation}
Therefore, if $\mathscr{R}^{\,2}=R^2$ then $R_1=R_2=R_3$, so the geometry
is isotropic.

\begin{figure}[!h]
{\centering
\includegraphics[height=0.32\textwidth]{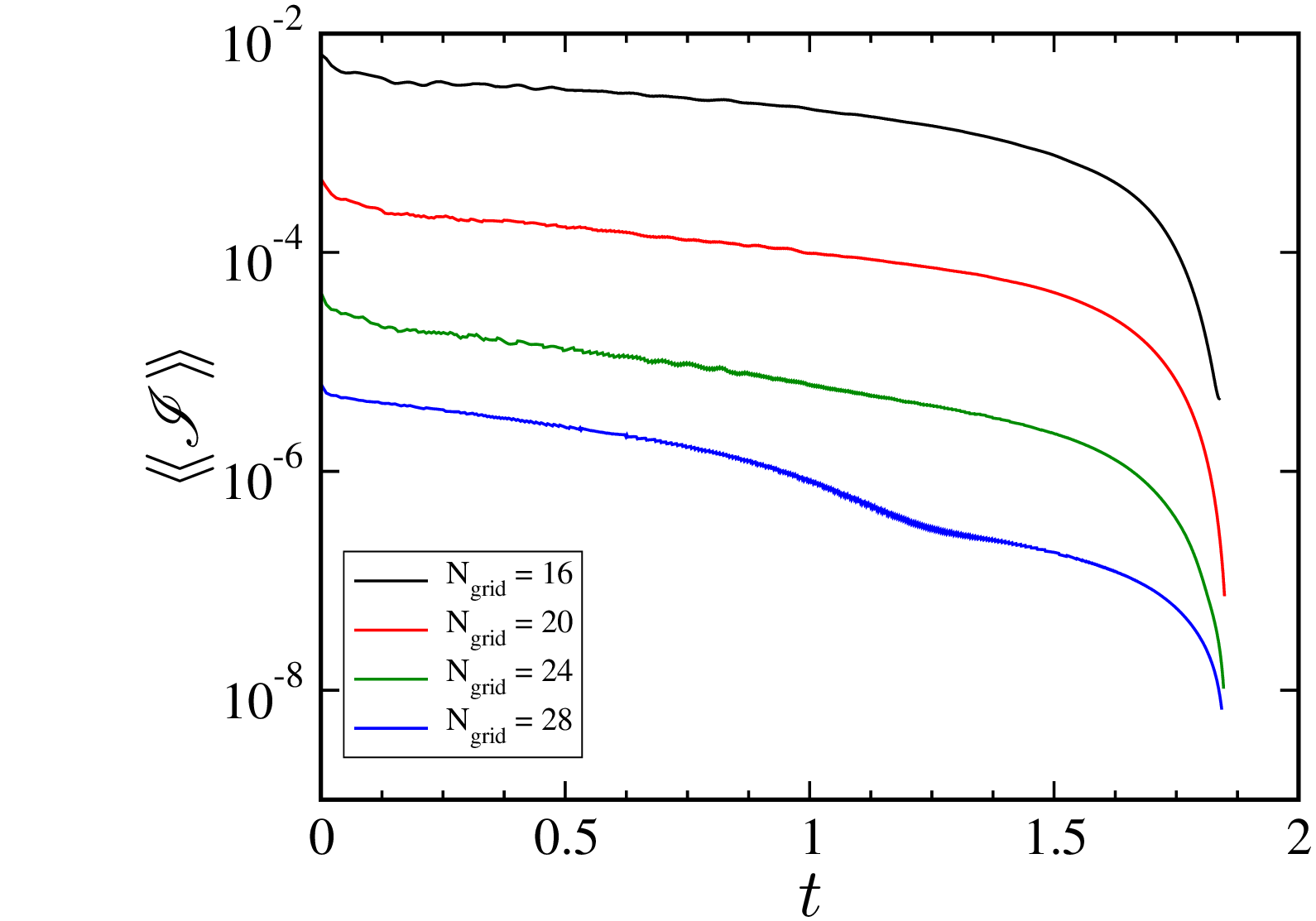}}
\caption{\label{f:E5_Isotrophy} This figure illustrates the isotropy
  parameter $\llangle\mathscr{I}\rrangle$ defined in
  Eq.~(\ref{e:IsotropyMeasure}) for evolutions on the Sixth-Turn
  manifold.}
\end{figure}

\begin{figure}[!h]
{\centering
\includegraphics[height=0.32\textwidth]{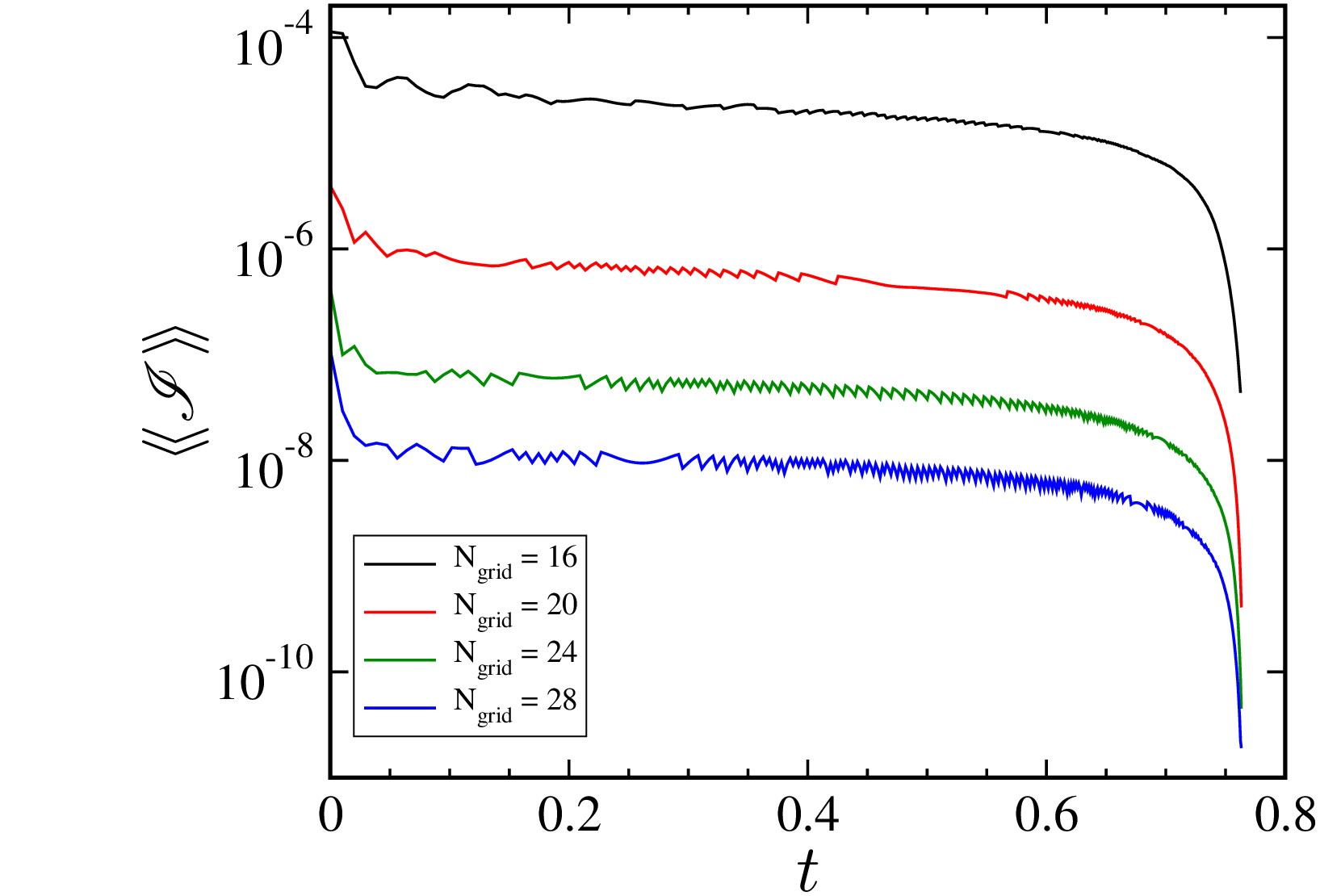}}
\caption{\label{f:L52_Isotropy} This figure illustrates the isotropy
  parameter $\llangle\mathscr{I}\rrangle$ defined in
  Eq.~(\ref{e:IsotropyMeasure}) for evolutions on the $L(5,2)$
  manifold.}
\end{figure}

\begin{figure}[!h]
{\centering
\includegraphics[height=0.32\textwidth]{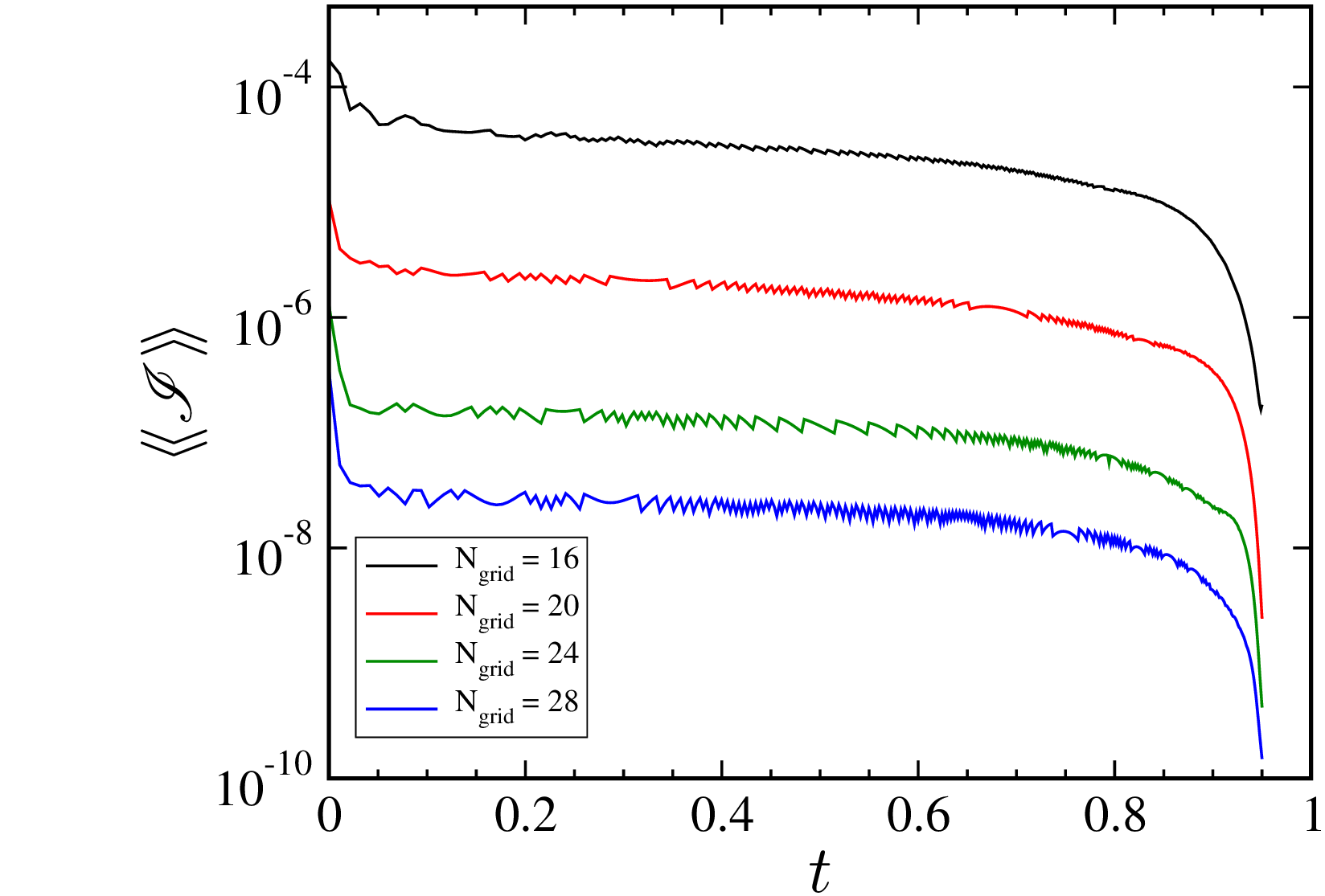}}
\caption{\label{f:SFS_Isotrophy} This figure illustrates the
  isotropy parameter $\llangle \mathscr{I}\rrangle$ defined in
  Eq.~(\ref{e:IsotropyMeasure}) for evolutions on the
  SFS[S2:(2,1)(2,1)(2,-1)] manifold.}
\end{figure}

One practical measure of the isotropy of the geometry is given by the
quantity $\llangle \mathscr{I}\rrangle$ defined by
\begin{equation}
  \llangle\mathscr{I}\rrangle
  =\bigl| |\llangle R\,\rrangle | -\llangle\mathscr{R}\,\rrangle\bigr|,
  \label{e:IsotropyMeasure}
\end{equation}
where $\llangle R \rrangle$ is defined in Eq.~(\ref{e:RicciScalarAve})
and $\llangle\mathscr{R}\,\rrangle$ is defined by
\begin{eqnarray}
  \llangle\mathscr{R}\,\rrangle &=& \frac{1}{\mathcal{V}}\int
  \mathscr{R}\,\sqrt{g}\,d^{\,3}x.
  \label{e:CurvatureScalarAve}
\end{eqnarray}
Figures~\ref{f:E5_Isotrophy}--\ref{f:SFS_Isotrophy} illustrate
$\llangle \mathscr{I}(t)\rrangle$ for the evolutions performed for
this study.  These results show that these cosmological models
are isotropic, in addition to being homogeneous.
%

%%%%%%%%%%%%%%%%%%%%%%%%%%%%%%%%%%%%%%%%%%%%%%%%%%%%%%%%%%%%%%%%%%%%%%%%%%%%%%%
\section{Discussion}
\label{s:Discussion}
%%%%%%%%%%%%%%%%%%%%%%%%%%%%%%%%%%%%%%%%%%%%%%%%%%%%%%%%%%%%%%%%%%%%%%%%%%%%%%%

Previous numerical studies~\cite{Zhang2026a, Zhang2026b, Zhang2026c}
of cosmological models on manifolds with non-trivial topologies used
the simple harmonic gauge, $H_a=0$, to determine the evolution of the
spacetime coordinates.  The lapse $N(t)$ in those cosmological models
evolved as $N(t)\backsimeq a^3(t)$, where $a(t)$ is the scale factor.
This made it impossible to reliably evolve large expansion models
(i.e. models with $a(t)\gg 1$) because the spacetime metric becomes
ill conditioned and therefore difficult to invert numerically.  This
study developed a new generalized harmonic gauge condition in
Sec.~\ref{s:CosmoGaugeCondition}, Eqs.~(\ref{e:NewGaugeDef}) and
(\ref{e:NewHkGaugeDef}), designed to make the growth of the lapse
$N(t)$ commensurate with the growth of the scale factor $a(t)$:
$N(t)\backsimeq a(t)$.  The tests reported in
Sec.~\ref{s:CosmoGaugeCondition} confirm that the new gauge condition
is successful in keeping the evolution of the lapse roughly
proportional to the evolution of the scale factor.

Additional tests of this new gauge were reported in
Sec.~\ref{s:LargeExpansionCosmologies} that evolved homogeneous and
isotropic initial data on exemplar manifolds with three different
spatial topologies: Sixth-Turn space, the lens space $L(5,2)$, and the
Seifert fiber space $SFS[S2:(2,1)(2,1)(2,-1)]$.  These tests show that
cosmological models can be evolved reliably using this new gauge long
enough for the spatial volumes to increase by a factor of a few million,
which is a significant increase over the factor of ten increase
achieved using harmonic gauge in Refs.~\cite{Zhang2026a, Zhang2026b,
  Zhang2026c}.

While the new gauge condition presented in
Sec.~\ref{s:CosmoGaugeCondition} makes it possible to perform
numerical cosmological evolutions that expand the spatial volume by a
factor of a few million, it is natural to ask why these evolutions
could not be extended to factors of billions or trillions.  It seems
likely that the current limitation is caused by errors in the time
stepping process of the numerical evolutions.  The parameters of the
Dormand-Prince time stepping algorithm used in this study were set to
achieve error tolerances at the $10^{-7.5}$ level, but to limit the
minimum time stepsize to $\min \delta t = 5\times 10^{-8}$.  The
evolution of the lapse $N(t)$ in this new gauge grows rapidly, so the
proper time increment $\delta \tau=N(t)\delta t$ becomes very large
and is increasing rapidly at the end of the evolutions presented in
Sec.~\ref{s:LargeExpansionCosmologies} where $\delta\tau \backsimeq
200 \delta t$.  The minimum time step limitation was enforced at the
ends of some of these evolutions.  Therefore it seems likely that the
adaptivity of the time stepping algorithm was not able preserve the
needed accuracy under these conditions.

If the time stepping errors as described above are the primary factor
limiting these evolutions, it might be possible to correct this
problem by a simple application of the dual-frame evolution algorithm
developed in Ref.~\cite{Scheel2006} for binary black hole evolutions.
In this approach the spacetime coordinates determined by the
generalized harmonic gauge developed in
Sec.~\ref{s:CosmoGaugeCondition} are used as the tensor bases in which
to represent the components of the various dynamical fields:
$u^\alpha=\{\psi_{ab},\Pi_{ab},\Phi_{iab}\}$ .  The standard
representation of the Einstein evolution equations using these
coordinates can be written (symbolically) in the form
\begin{equation}
  \partial_t u^\alpha =- A^{k\alpha}{}_\beta({\bf
    u})\tilde\nabla_ku^\beta +F^\alpha({\bf u},{\bf \tilde\Gamma},
          {\bf \partial\tilde\Gamma})
  \label{e:EvolutionEqs0}
\end{equation}
(see Appendix~\ref{s:SymmetricHyperbolicSystem}).  This standard
representation of the equations is the one used compute the models
reported in Sec.~\ref{s:LargeExpansionCosmologies}, i.e. the models
that have the time stepping problem outlined above.  The equations in
this form determine the dynamical fields as functions of the original
spacetime coordinates: $u^\alpha=u^\alpha(t,x^i)$.  A simple version
of the dual frame idea is to change the time coordinate used in the
numerical evolution of these equations, without changing the time
coordinate used in the representation of the various components of the
dynamical fields.  Let $s$ denote this new time coordinate defined by
the relationship $ds = f(t)dt$.  Using this new coordinate the
Einstein evolution equations become
\begin{equation}
  \partial_s u^\alpha =f^{-1}(s)\left[- A^{k\alpha}{}_\beta({\bf
    u})\tilde\nabla_ku^\beta +F^\alpha({\bf u},{\bf \tilde\Gamma},
          {\bf \partial\tilde\Gamma})\right].
  \label{e:EvolutionEqs0}
\end{equation}
The numerically determined solutions to this equation for the
dynamical fields have the form $u^\alpha=u^\alpha(s,x^i)$.  Choosing
$f(s)=N(s)$ would make $s$ equal to the physical proper time of
comoving observers in the $t=$ constant hypersurfaces.  Integrating
the equations using proper time should avoid the type of time stepping
problem described above that is caused by the rapidly growing lapse.
It is also possible that the choice $f(s)=N(s)$ would make the time
integrations so slow and inefficient that it would not be possible
practically to achieve large spatial expansions.  So it is possible
that a more general time re-scaling, e.g. $f(s)=N^\nu(s)$ for some
constant $\nu<1$, would be better.  We plan to explore these
possibilities if and when performing cosmological evolutions with even
larger spatial expansions are needed.

\appendix
%%%%%%%%%%%%%%%%%%%%%%%%%%%%%%%%%%%%%%%%%%%%%%%%%%%%%%%%%%%%%%%%%%%%%%%%%%%%%%%
\section{Covariant First-Order Symmetric-Hyperbolic Einstein System}
\label{s:SymmetricHyperbolicSystem}
%%%%%%%%%%%%%%%%%%%%%%%%%%%%%%%%%%%%%%%%%%%%%%%%%%%%%%%%%%%%%%%%%%%%%%%%%%%%%%%
\renewcommand{\theequation}{A.\arabic{equation}} % Redefines equation numbering
\setcounter{equation}{0} % Resets the equation counter for the appendix

This study uses a covariant representation of the Einstein equations
that enforces generalized harmonic gauge
conditions~\cite{Lindblom2014}.  The covariance of this representation
is essential for the study of solutions on manifolds with multiple
multi-cube coordinate patches, such as the manifolds studied here.
All the dynamical fields in this representation are tensors, so they
can be transformed in a straightforward way across the interfaces
between coordinate patches.  The dynamical fields, $u^\alpha$, for the
first-order symmetric-hyperbolic representation of the Einstein
equations used in this study are the collection of tensor fields,
\begin{equation}
  u^\alpha = \left\{ \psi_{ab}, \Pi_{ab}, \Phi_{iab}\right\},
  \label{e:UalphaDef}
\end{equation}
where $\psi_{ab}$ is the spacetime metric, and the tensors $\Pi_{ab}$
and $\Phi_{iab}$ represent its first derivatives,
\begin{eqnarray}
  \Pi_{ab} &=& -t^c\tilde\nabla_c\psi_{ab},
  \label{e:PiDef}\\
  \Phi_{iab}&=& \tilde\nabla_i\psi_{ab}. 
  \label{e:PhiDef}
\end{eqnarray} 
The derivative $\tilde\nabla_c$ used in these expressions is the
torsion-free covariant derivative compatible with the reference metric
$\tilde\psi_{ab}$: $\tilde\nabla_c\tilde\psi_{ab}=0$.  This reference
metric is also used to define the Jacobians used to transform tensor
fields across the boundary interfaces between multi-cube regions.  The
vector $t^c$ represents the unit timelike normal ($t^a t^b
\psi_{ab}=-1$) to the spacelike hypersurfaces used to perform the
evolutions.  The lower range of Latin indices, e.g. $a$, $b$, $c$,
..., range over the four spacetime coordinates, while the higher range
of indices, e.g. $i$, $j$, $k$, ..., range over the three spatial
coordinates on each $t=$ constant hypersurface.  The Greek indices,
e.g. $\alpha$, $\beta$, $\gamma$, range over the fifty components of
the dynamical tensor fields $u^\alpha$.  The Einstein equations for
these fields can be written in the first-order form
\begin{equation}
  \partial_t u^\alpha
  + A^{k\alpha}{}_\beta({\bf u})\tilde\nabla_ku^\beta
  =F^\alpha({\bf u},{\bf \tilde\Gamma}, {\bf \partial\tilde\Gamma}),
  \label{e:EvolutionEqs}
\end{equation}
where the tensors $ A^{k\alpha}{}_\beta({\bf u})$ and $F^\alpha({\bf
  u},{\bf \tilde\Gamma}, {\bf \partial\tilde\Gamma})$ depend in
complicated ways on the dynamical fields $u^\alpha$, the reference
connection $\tilde\Gamma^c{}_{ab}$ and its derivatives
$\partial_d\tilde\Gamma^c{}_{ab}$.  The detailed expressions for this
representation of the covariant Einstein evolution system are given in
Ref.~\cite{Lindblom2014}.

The spacetime coordinates are determined in the generalized harmonic
representations of Einstein's equation by imposing a condition on the
difference between the connection $\tilde\Gamma^c{}_{ab}$ associated
with the reference metric $\tilde \psi_{ab}$ and the physical
connection $\Gamma^c{}_{ab}$ associated with the physical spacetime
metric $\psi_{ab}$.  In particular the difference between these
connections is fixed by a gauge source function $H_a$:
\begin{equation}
H_a =  -\psi_{ad}\,\psi^{bc}\left(\Gamma^d{}_{bc}-\tilde\Gamma^d{}_{bc}\right).
  \label{e:HarmonicGageDef}
\end{equation}
(Note that the difference between any two connections is a tensor.)
In general $H_a$ may be any function that depends on the physical
metric $\psi_{ab}$ (but not its derivatives) and the reference metric
$\tilde\psi_{ab}$.  The gauge condition used for the numerical
evolutions performed in this study is the one given in
Eqs.~(\ref{e:NewGaugeDef}) and (\ref{e:NewHkGaugeDef}):
$t^aH_a=\mu\log(N/a^\sigma)$ and $H_k=0$.

The generalized harmonic evolution system contains a number of
constraints.  In particular the gauge condition,
Eq.~(\ref{e:HarmonicGageDef}) is in effect a constraint
\begin{equation}
  \mathcal{C}_a =H_a +
  \psi_{ad}\psi^{bc}\left(\Gamma^d{}_{bc}-\tilde\Gamma^d{}_{bc}\right).
  \label{e:C1Def}
\end{equation}
on the dynamical fields.  In addition, the equation that defines
$\Phi_{iab}$, Eq. (\ref{e:PhiDef}), is also a constraint
\begin{equation}
  \mathcal{C}_{iab} = \tilde\nabla_i\psi_{ab}-\Phi_{iab}.
  \label{e:C3Def}
\end{equation}
These primary constraints, $\mathcal{C}_a$ and $\mathcal{C}_{iab}$,
satisfy a second-order system of evolution equations as a consequence
of the Einstein evolution system, Eq.~(\ref{e:EvolutionEqs}). (See
Ref.~\cite{Lindblom2014} for details.)  This second-order constraint
evolution system can be converted to a first-order symmetric
hyperbolic system by introducing the secondary constraints,
\begin{eqnarray}
  \mathcal{F}_a &=& t^c\nabla_c\mathcal{C}_a,
  \label{e:F1Def}\\
  \mathcal{C}_{ia} &=& \nabla_i\mathcal{C}_a,
  \label{e:C2Def}\\
  \mathcal{C}_{ijab}&=&2\tilde\nabla_{[i}\mathcal{C}_{j]ab}.
  \label{e:C4Def}
\end{eqnarray}
The symmetric-hyperbolic evolution equation for the collection of
constraints,
\begin{equation}
  \mathcal{C}^\alpha=\left\{\mathcal{C}_a,\mathcal{C}_{iab},
  \mathcal{F}_a,\mathcal{C}_{ia},\mathcal{C}_{ijab}\right\},
  \label{e:ConstraintTotal}
\end{equation}
ensures that solutions to the Einstein Eq.~(\ref{e:EvolutionEqs}) that
satisfy the constraints, $\mathcal{C}^\alpha$, on an initial surface
will satisfy them throughout the evolution of those initial data.
Numerical solutions to the equations will of course contain (hopefully
small) violations of these constraints.  This study monitors the size
of these constraint violations by evaluating a norm constructed from
the composite constraint, $\mathcal{C}_\psi$, defined by,
\begin{eqnarray}
  \mathcal{C}_\psi^2 &=& \delta^{ab}\left(\mathcal{C}_a\mathcal{C}_b
  +\mathcal{F}_a\mathcal{F}_b\right)\nonumber\\
&&  +\tilde g^{ij}\delta^{ab}\delta^{cd}\left(\mathcal{C}_{iac}\mathcal{C}_{jbd}
    +\tfrac{1}{4}\tilde g^{kl}\mathcal{C}_{ikac}\mathcal{C}_{jlbd}\right).
    \qquad
  \label{e:CpsiDef}
\end{eqnarray}
This study evaluates a norm of $\mathcal{C}_\psi$,
$\|\,\mathcal{C}_\psi\,\|$, defined by
\begin{equation}
  \|\,\mathcal{C}_\psi\,\|^2 = \frac{\int \mathcal{C}_\psi^2\sqrt{g}\,d^{\,3}x}
  {\int\mathcal{N}_\psi^2\sqrt{g}\,d^{\,3}x}.
  \label{e:CpsiNormDef}
\end{equation}
The quantity that appears in the denominator, $\mathcal{N}_\psi$,
is defined by
\begin{eqnarray}
  \mathcal{N}_\psi^2&=& \delta^{ab}\delta^{cd}\tilde g^{ij}\left(
  \partial_i\psi_{ac}\partial_j\psi_{bd}
  +  \partial_i\Pi_{ac}\partial_j\Pi_{bd}\right)\nonumber\\
  &&+\delta^{ab}\delta^{cd}\tilde g^{ij}
  \tilde g^{kl}\partial_i\Phi_{kac}\partial_j\Phi_{lbd}.
  \label{e:NpsiDef}
\end{eqnarray}
It has been included in the definition of $\|\,\mathcal{C}_\psi\,\|$ to 
provide a dimensionless measure of the fractional errors due to
constraint violations in the numerical solutions to the Einstein
evolution system.

%%%%%%%%%%%%%%%%%%%%%%%%%%%%%%%%%%%%%%%%%%%%%%%%%%%%%%%%%%%%%%%%%%%%%%%%%%%%%%%
% Initial Gauge Constraint on the Initial Hypersurface
%%%%%%%%%%%%%%%%%%%%%%%%%%%%%%%%%%%%%%%%%%%%%%%%%%%%%%%%%%%%%%%%%%%%%%%%%%%%%%%
\section{Gauge Constraint on the Initial Hypersurface}
\label{s:InitialGauge}
%%%%%%%%%%%%%%%%%%%%%%%%%%%%%%%%%%%%%%%%%%%%%%%%%%%%%%%%%%%%%%%%%%%%%%%%%%%%%%%
\renewcommand{\theequation}{B.\arabic{equation}} % Redefines equation numbering
\setcounter{equation}{0} % Resets the equation counter for the appendix

The gauge source function $H_a$ is used to fix the difference between
the physical connection $\Gamma^a_{bc}$ determined by the physical
metric $\psi_{ab}$, and the reference connection $\tilde
\Gamma^a_{bc}$ determined by the reference metric
$\tilde\psi_{ab}$. This gauge constraint can be written as,
\begin{equation}
  H_a=-\psi_{ad}\psi^{bc}\Bigl(\Gamma^d_{bc}-\tilde\Gamma^d_{bc}\Bigr).
  \label{e:GaugeConstraint}
\end{equation}
This Appendix gives a more complete description of how this constraint
influences the initial values of the $\Pi_{ta}$ components of the
first order dynamical field.

The reference metric used in this study is determined from the spatial
metric, $\tilde g_{ij}=g^0_{ij}$, determined from the solution to the
initial value equations on the $t=0$ hypersurface, and the assumption
that the reference metric lapse has the value $\tilde N=1$ and the
shift $\tilde N_i=0$.  The reference metric  therefore has the form,
\begin{equation}
  \tilde\psi_{ab}\,dx^adx^b=-dt^2+g^0_{ij}\,dx^idx^j.
  \label{e:ReferenceMetricDef}
\end{equation}
and is assumed to be time independent: $\partial_t\tilde\psi_{ab}=0$.

The initial data for the evolutions used in this study sets the
physical metric equal to the reference metric at $t=0$:
$\psi_{ab}=\tilde\psi_{ab}$.  Given this initial data, the expression
for the gauge constraint, Eq.~(\ref{e:GaugeConstraint}), can be
written more explicitly on the $t=0$ hypersurface:
\begin{eqnarray}
  H_a&=&-\tfrac{1}{2}\psi^{bc}\Bigl[
    \partial_b(\psi_{ac}-\tilde\psi_{ac})
    +\partial_c(\psi_{ab}-\tilde\psi_{ab})\nonumber\\
    &&\qquad\quad-\partial_a(\psi_{bc}-\tilde\psi_{bc})\Bigr].
\end{eqnarray}
On the $t=0$ hypersurface $\psi_{ab}=\tilde\psi_{ab}$ so the spatial
derivative terms, $\partial_k(\psi_{ab}-\tilde\psi_{ab})$, of the
reference metric vanish.  In addition, the time independent form of
the reference metric in Eq.~(\ref{e:ReferenceMetricDef}) implies that
all the derivatives $\partial_b\tilde\psi_{ta}$ of the temporal
components of the reference metric vanish.  Therefore the expression
for the gauge constraint can be simplified further on the $t=0$
hypersurface:
\begin{eqnarray}
  H_t&=& \tfrac{1}{2}\partial_t\psi_{tt}-K,\label{e:HtInitialData}\\
  H_k&=& \partial_t\psi_{kt}.\label{e:HkInitialData}
\end{eqnarray}

The dynamical field $\Pi_{ab}=-t^c\tilde\nabla_c\psi_{ab}$ on the
$t=0$ initial hypersurface has the simple form,
\begin{eqnarray}
  \Pi_{ab}=-\partial_t\psi_{ab},
  \label{e:PiabInial}
\end{eqnarray}
because $t^a\partial_a=\partial_t$ and $\partial_t\tilde\psi_{ab}=0$
on that surface.  Equations~(\ref{e:HtInitialData}) and
(\ref{e:HkInitialData}) show that the initial values of the $\Pi_{ta}$
components are determined by the initial value of the gauge source
vector on that hypersurface:
\begin{eqnarray}
  \Pi_{tt}&=& -2(H_t+K),\label{e:InitialPitt}\\
  \Pi_{tk}&=& -H_k.\label{e:initialPitk}
\end{eqnarray}
The generalized harmonic gauge source vector used in this study,
Eq.~(\ref{e:NewGaugeDef}), vanishes since $N(0)=a(0)=1$ on the $t=0$
initial hypersurface.  Therefore the initial values of the dynamical
field $\Pi_{ta}$ for the evolutions performed for this study have the
values,
\begin{eqnarray}
  \Pi_{tt}&=& -2K,\label{e:InitialPitt1}\\
  \Pi_{tk}&=& 0.\label{e:initialPitk1}
\end{eqnarray}
\vspace{0.1cm}

%%%%%%%%%%%%%%%%%%%%%%%%%%%%%%%%%%%%%%%%%%%%%%%%%%%%%%%%%%%%%%%%%%%%%%%%%%%%%%%
\begin{acknowledgements}
  We thank Michael Holst for providing the computational resources
  used to perform the numerical calculations reported in this study.
  L.L. was supported in part by grant No. 2407545 from the National
  Science Foundation to the University of California at San Diego,
  USA. F.Z. was supported by the National Key Research and Development
  Program of China grant 2023YFC2205801, and the National Natural
  Science Foundation of China grants 12433001 and 12021003.
\end{acknowledgements}
%%%%%%%%%%%%%%%%%%%%%%%%%%%%%%%%%%%%%%%%%%%%%%%%%%%%%%%%%%%%%%%%%%%%%%%%%%%%%%%

%%%%%%%%%%%%%%%%%%%%%%%%%%%%%%%%%%%%%%%%%%%%%%%%%%%%%%%%%%%%%%%%%%%%%%%%%%%%%%%
% References
%%%%%%%%%%%%%%%%%%%%%%%%%%%%%%%%%%%%%%%%%%%%%%%%%%%%%%%%%%%%%%%%%%%%%%%%%%%%%%%
%\clearpage
\bibliography{../References/References}

%%%%%%%%%%%%%%%%%%%%%%%%%%%%%%%%%%%%%%%%%%%%%%%%%%%%%%%%%%%%%%%%%%%%%%%%%%%%%%%
\end{document}